\documentclass[reprint, 
amsmath,
amssymb,
]{revtex4-1}
\usepackage{braket}
\usepackage{amsmath}
\usepackage[colorlinks,linkcolor=red,anchorcolor=blue,citecolor=blue]{hyperref}
\usepackage{graphicx}
\usepackage{epstopdf}
\usepackage{dcolumn}
\usepackage{bm}
\usepackage{verbatim}
\usepackage{amsfonts}
\usepackage{color}

\begin{document}

\preprint{APS/123-QED}

\title{Strong unitary uncertainty relations}

\author{Bing Yu} 
\affiliation{School of Mathematics, South China University of Technology, Guangzhou 510640, China}
\author{Naihuan Jing}
\email{jing@ncsu.edu}
\affiliation{School of Mathematics, South China University of Technology, Guangzhou 510640, China}
\affiliation{Department of Mathematics, North Carolina State University, Raleigh, NC 27695, USA}
\author{Xianqing Li-Jost}
\affiliation{Max-Planck-Institute for Mathematics in the Sciences, 04103 Leipzig, Germany}
\date{August 13, 2019}

\begin{abstract}

In this paper we provide a new set of
uncertainty principles for unitary operators using a sequence of inequalities with the help of the geometric-arithmetic mean inequality. As these inequalities are ``fine-grained'' compared with the well-known Cauchy-Schwarz inequality,
our framework naturally improves the results based on the latter. As such, the unitary uncertainty relations based on our method outperform the best known bound introduced in [Phys. Rev. Lett. 120, 230402 (2018)] to some extent. Explicit examples of unitary uncertainty relations are provided to back our claims.

\end{abstract}

\maketitle


\section{\label{sec:leve1}Introduction}
At the foundation of quantum theory lies the Heisenberg uncertainty principle \cite{Heisenberg1927}, which was first introduced in 1927. Traditionally, the textbook version of the uncertainty relation was established by Kennard \cite{Kennard1927} (see also the work of Weyl \cite{Weyl1927}) by means of variance in terms of position and momentum.
The uncertainty principle lets us understand that if we were able to measure the momentum of a quantum system with certainty, then we would not gain the information of the measurement outcome of location with certainty. Robertson \cite{Robertson1929} generalized the uncertainty relation for position and momentum to any two bounded observables $A$ and $B$ as
\begin{equation}
\label{Robertson}
\Delta A \Delta B \geqslant \frac{1}{2} | \langle\psi| [A,B] |\psi\rangle |,
\end{equation}
where $\Delta$ stands for the standard deviation of the observable relative to a fixed state $\ket{\psi}$ and $[A,B]$ represents the commutator of the observables $A$ and $B$. Later Eq. (\ref{Robertson}) was improved by Schr\"odinger \cite{Schrodinger1930}. Recently, variance-based uncertainty relations have been intensely studied in \cite{Massar2008,Klimov2009,Marchiolli2012,Namiki2012,Marchiolli2013,Maccone2014,Li2015,Pati2015,Hall2016,Bagchi2016,
Xiao2016W,Xiao2016S,Mondal2017,Xiao2017I,XiaoL2017,Qian2018,Sazim2018,Sharma2018,Bong2018}.

Because of their relevance in quantum information theory, the entropies \cite{Bialynicki1975,Deutsch1983,Partovi1983,Kraus1987,Maassen1988,Ivanovic1992,Sanchez1993,Ballester2007,Wu2009,Huang2011,
Tomamichel2011,Coles2012,Coles2014,Xiao2016SE,Xiao2016QM, Xiao2016U, Coles2017} have been employed to quantify the uncertainty relations between incompatible observables. 
The entropies 
are by no reason the best way to formulate joint uncertainties, and it is reasonable to consider all nonnegative Schur-concave functions as qualified uncertainty measures. This has lead to the well known universal uncertainty relations \cite{Friedland2013,Puchala2013,Rudnicki2014,Narasimhachar2016}
expressed by majorization \cite{Majorization}. To this end, we shall remark that all these uncertainty relations play an important role in a wide range of applications such as 
entanglement detection \cite{Guhne2004,Hofmann2003}, quantum spin squeezing \cite{Walls1981,Wodkiewicz1985,Wineland1992,Kitagawa1993,Ma2011},  quantum metrology \cite{Braunstein1994,Braunstein1996,Giovannetti2004,Giovannetti2006,Giovannetti2011}, quantum nonlocality \cite{Oppenheim2010,Xiao2018} and so on.

Now we turn to the variance-based uncertainty relations in the product form for unitary operators.
Massar and Spiandel \cite{Massar2008} have considered the uncertainty relation for two unitary operators that
satisfy the commutation relation $UV=e^{i\phi}VU$. This uncertainty relation gives rise to the constraint for
a quantum state to be simultaneously localized in two mutually unbaised bases 
related by a discrete Fourier transform (DFT). Other applications of Masser-Spiandel's uncertainty relations include
modular variables \cite{Aharonov1969} and signal processing \cite{Opatrny1995,Opatrny1996}. 
Several further uncertainty relations for unitary operators related by DFT
have been investigated in \cite{Marchiolli2012,Marchiolli2013,Klimov2009,Namiki2012}. Later Bagchi and Pati \cite{Bagchi2016} derived sum-form variance-based uncertainty relations for two general unitary operators, which have been tested experimentally with photonic qutrits \cite{XiaoL2017}. The uncertainty relation for two general unitary operators is directly related to the preparation uncertainty principle that the amount of visibility for noncommuting unitary operators is nontrivially upper bounded.
It is noted that 
 a crucial technique underlying the variance-based uncertainty relations for two observables or unitary operators is the celebrated Cauchy-Schwarz inequality.

For multi-observables, the generalized uncertainty relation was first considered by Robertson using the positive semidefiniteness of a Hermitian matrix \cite{Robertson1934}.
Recently, Bong et al used a similar method to derive a strong variance-based uncertainty relation for any $n$ unitary operators \cite{Bong2018}.
The unitary uncertainty relation
implies 
the famous Robertson-Sch\"{o}dinger uncertainty relation in the case of two Hermitian operators \cite{Schrodinger1930,Robertson1934}.
However, the lower bound is implicitly given and sometimes hard to
compute. This raises the question of explicitly extracting the uncertainty relation from the Gram determinant and also one wonders whether this 
strong
uncertainty relation can be further improved.

The goal of this paper is to give new and improved uncertainty relations for general unitary operators.
Following Xiao et al \cite{Xiao2016S}, a sequence of ``fine-grained'' inequalities compared with
the Cauchy-Scharz inequality are employed to derive uncertainty relations in connection with the Geometric-Arithmetic mean (AGM) inequality.
We use this method to derive new variance-based unitary uncertainty relations in the product form for two and three operators
 in all quantum systems.
The new uncertainty bounds for two unitary operators outperform those of Bong et al's in the whole range. As the improvement is due to replacement of the Cauchy-Schwarz inequality underlying all previous uncertainty principles, our method provides fundamentally better bounds.
We also generalize the uncertainty relation to the case of multiple unitary operators, and the new lower bounds are also shown to be tighter
than that of Bong et al's to some extent.

This paper is organized as follows. In Sec.\ref{sec:leve2} we introduce a fine-grained sequence of inequalities to generalize the Cauchy-Schwarz inequality, which was proved twice in this consideration. Our first main result (Thm. 1) of variance-based unitary uncertainty relations in the product-form is given in Sec.\ref{sec:leve21} for two unitary operators.
In Sec.\ref{sec:leve22}, the bounds are strengthened by symmetry of permutations. In Sec.\ref{sec:leve23}, examples are given to show our Theorem.1 provides tighter bounds than those of Bong et al's. In Sec.\ref{sec:leve3}, we investigate product-form variance-based unitary uncertainty relations for three unitary operators. The uncertainty relations for multiple unitary operators are addressed in Sec.\ref{sec:leve31}, and comparison is also provided with previous lower bounds for
qutrit pure state, four-dimensional pure state and qutrit mixed state are studied in Sec.\ref{sec:leve32}.
Concluding remarks are given in \ref{sec:leve4}. In the Appendix (Sec. \ref{appendix}), we give some details of the proofs and calculations.

\section{\label{sec:leve2}Uncertainty Relations for two unitary operators}
Let $A$ and $B$ be two unitary operators defined in a finite-dimensional Hilbert space with a fixed state $\ket{\psi}$. With respect to the mean value $\langle A\rangle=\bra{\psi}A\ket{\psi}$, the variance of $A$ over $\ket{\psi}$ is defined by
\begin{align}
\Delta A^2&=\langle(A-\langle A\rangle)^{\dag}(A-\langle A\rangle)\rangle\notag\\
&=\bra{\psi}\delta\hat{A}^\dag\delta\hat{A}\ket{\psi}
\end{align}
where $\delta\hat{A}=A-\langle A\rangle$.
Note that the variance is bounded by $0\leqslant\Delta A^2\leqslant1$.

Suppose $\{\ket{\psi_1}, \cdots, \ket{\psi_n}\}$ is a computational basis, then the state
$\ket f=\delta\hat{A}\ket{\psi}$ can be written as $\ket f=\sum_{j=1}^n\alpha_j\ket{\psi_j}$
and similarly $\ket g=\delta\hat{B}\ket{\psi}=\sum_{j=1}^n\beta_j\ket{\psi_j}$.
Thus the product of the variances obeys the unitary uncertainty relation (UUR)
\begin{align}\label{prod2}
&\Delta A^2\Delta B^2=\langle f|f\rangle\langle g|g\rangle\notag=\sum\limits_{i,j}|\alpha_i|^2|\beta_j|^2\notag\\
&\geqslant|\sum_{i=1}^n\alpha_i^\ast\beta_i|^2=|\langle f|g\rangle|^2 \\
&=|\langle A^\dag B\rangle-\langle A^\dag\rangle\langle B\rangle|^2\notag,
\end{align}
where the inequality is due to the Cauchy-Schwarz inequality. Note that the last expression is independent from
the choice of the computational basis.

Let $\overrightarrow{X}=(x_1, x_2, \cdots, x_n)$ and $\overrightarrow{Y}=(y_1, y_2, \cdots, y_n)$ be the (nonnegative) real vectors
given by $x_i=|\alpha_i|$, $y_j=|\beta_j|$, where $(\alpha_1, \ldots, \alpha_n)$ and $(\beta_1, \ldots, \beta_n)$ are the coordinate vectors
of $\delta\hat{A}$ and $\delta\hat{B}$ respectively.
Then the product of the variances 
can be rewritten as $\Delta A^2\Delta B^2=|\overrightarrow{X}|^2|\overrightarrow{Y}|^2=\sum\limits_{i,j}x_i^2y_j^2$.
Note that 
the Cauchy-Schwarz inequality is in fact a consequence of
$n(n-1)/2$ AGM inequalities. Indeed,
\begin{align}\label{amgm}
\sum\limits_{i,j}x_i^2y_j^2&=\sum\limits_{i<j}(x_i^2y_j^2+x_j^2y_i^2)+\sum\limits_{i}x_i^2y_i^2\notag\\
&\geqslant\sum\limits_{i<j}2x_iy_jx_jy_i+\sum\limits_{i}x_i^2y_i^2\\
&=(\sum_{i=1}^n x_iy_i)^2\notag
\end{align}
with equality if and only if $x_iy_j=x_jy_i$ for all $i\neq j$.

\begin{figure}[ht]
  \centering
  \includegraphics[width=3.0in]{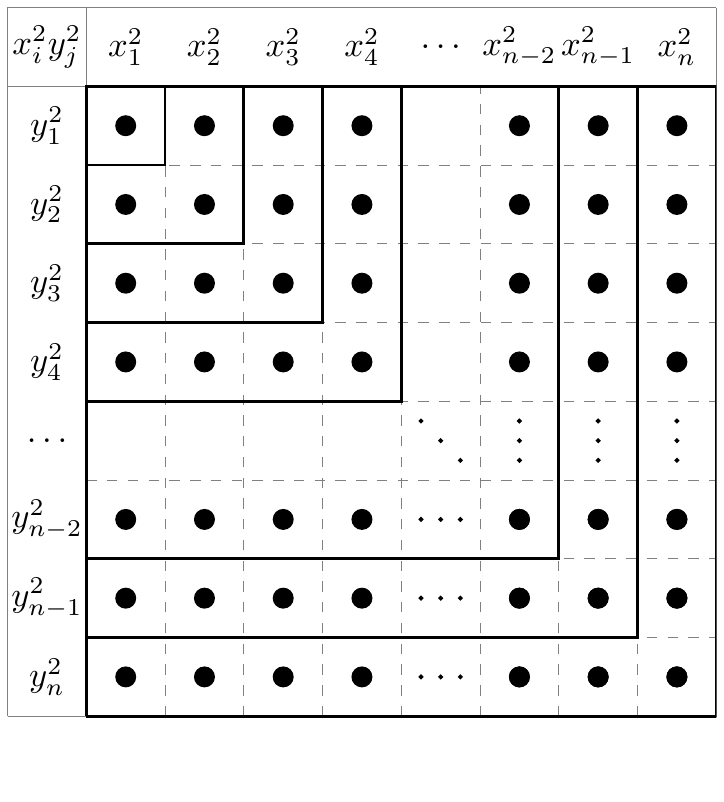}
  \caption{\textbf{Diagram for the $I_k$} ($1\leqslant k\leqslant n$).
  The black $(i, j)$-dot represents $x_i^2y_j^2$.
  So $I_k$ is $(\sum\limits_{i=1}^k x_iy_i)^2$ plus the dots outside of the $k$th principal square:
  $I_k=(\sum\limits_{i=1}^k x_iy_i)^2+\sum\limits_{1\leqslant i<j\leqslant n\atop k<j}(x_i^2y_j^2+x_j^2y_i^2)+\sum\limits_{k+1\leqslant i\leqslant n}x_i^2y_i^2$.
  The $k$th principal square shows the Cauchy-Schwarz inequality:
   $\sum\limits_{i,j=1}^k x_i^2y_j^2\geqslant(\sum\limits_{i=1}^k x_iy_i)^2$.}
  \label{hierarchy}
\end{figure}

Now we refine the Cauchy-Schwarz inequality
by introducing a sequence of partial ones.
For each $1\leq k\leq n$, define
\begin{align}\notag
I_k&=\sum\limits_{1\leq i\leq n}x_i^2y_i^2+\sum\limits_{\substack{1\leqslant i<j\leqslant n\\k<j}}(x_i^2y_j^2+x_j^2y_i^2)\\
&\qquad +\sum\limits_{1\leqslant i<j\leqslant k}2x_iy_ix_jy_j.
\end{align}
In particular,
$I_1=|\overrightarrow{X}|^2|\overrightarrow{Y}|^2$ and $I_n=(\sum_{i=1}^n x_iy_i)^2$. 
The quantities $I_k$ can be vidualized by lattice dots within an $n\times n$ square as follows. In Fig.\ref{hierarchy}
the black dot at $i$th column and $j$th row presents $x_i^2y_j^2$, then $I_k$ is the quantity
$(\sum\limits_{i=1}^k x_iy_i)^2$ plus the dots outside of the $k$th principal square. It is easily seen that
\begin{align*}
I_{k+1}-I_k
=-(\sum_{i=1}^{k}x_iy_{k+1}+y_ix_{k+1})^2\leq 0.
\end{align*}
One therefore obtains the following descending
sequence 
\begin{align}\label{Chain}
I_1\geqslant I_2\geqslant\cdots\geqslant I_{n-1}\geqslant I_n
\end{align}
and the Cauchy-Schwarz inequality also follows from the sequence: $I_1\geqslant I_n$.

\subsection{\label{sec:leve21}Main Results}

Let $\rho$ be a mixed state on the Hilbert space. The variance of the unitary operator $A$ with respect to
$\rho$ is defined as
\begin{align}
(\Delta A)^2=\mathrm{Tr}(\rho\delta\hat{A}^\dag\delta\hat{A})
\end{align}
Let $M=(m_{ij})_{l\times p}$ be a rectangular matrix, the vectorization $|M\rangle$ (or $vec(M)$) is the straightening vector $(m_{11}, \ldots, m_{1p}, \ldots, m_{l1}, \ldots, m_{lp})\in\mathbb{C}^{lp}$. As $\rho$ 
is positive semi-definite, we will
simply denote by $|\sqrt\rho\rangle$ the pure state given by the vectorization $vec(\sqrt\rho)$ in the computational basis. Note that the vector
$|\sqrt\rho\rangle$ satisfies the following property \cite{Dhrymes1978}
\begin{align}\label{vec}
\ket{MT}=(I\otimes M)\ket{T}
\end{align}
for two matrices $M$ and $T$ in suitable size. Thus
\begin{align}\label{variancea}
\Delta A^2
&=\mathrm{Tr}(\sqrt{\rho}\delta\hat{A}^\dag\delta\hat{A}\sqrt{\rho})\notag\\
&=\bra{\sqrt{\rho}}(I\otimes\delta\hat{A}^\dag\delta\hat{A})\ket{\sqrt{\rho}}\notag\\
&=|(I\otimes\delta\hat{A})\ket{\sqrt{\rho}}|^2,
\end{align}
where $\sqrt{\rho}$ is the uniquely defined semi-definite positive matrix associated to $\rho$.

\emph{Theorem 1.} Let $A$ and $B$ be two unitary operators on an $n$-dimensional Hilbert space $H$ and
$\rho$ a quantum state on $H$. Suppose $x_i$ and $y_i$ are the probabilities of
$\delta\hat{A}$ and $\delta\hat{B}$ with respect to a computational basis of $H$.
Then the product of the variances of $A$ and $B$ satisfies
the following uncertainty relations ($k=1, \ldots, N$)
\begin{equation}\label{r:1}
\Delta A^2\Delta B^2\geqslant I_k,
\end{equation}
where $N=n$ (or $n^2$) if $\rho$ is pure (or mixed),
$I_k=\sum\limits_{1\leqslant i\leqslant N}x_i^2y_i^2+\sum\limits_{\substack{1\leqslant i<j\leqslant N\\k<j}}(x_i^2y_j^2+x_j^2y_i^2)+\sum\limits_{1\leqslant i<j\leqslant k}2x_iy_jx_jy_i$ and the equality holds if and only if $x_iy_j=x_jy_i$ for all $1\leqslant i\neq j\leqslant k$.

\

\emph{Proof}. The uncertainty relations (\ref{r:1}) for the case of pure state $\rho$ were already shown in the last section.
As for the mixed state $\rho$, we remarked that $\ket{\sqrt\rho}$ is viewed as a pure state in an $n^2$ dimensional Hilbert space \cite{Watrous2011}, therefore
the relations (\ref{r:1}) also follow for all $k=1, \ldots, n^2$.

\emph{Remark 1}. Note that $|\overrightarrow{X}|^2|\overrightarrow{Y}|^2\geqslant I_k$ amounts to a partial Cauchy-Schwarz inequality \cite{Xiao2016S} as it is obtained by applying the Cauchy-Schwarz inequality on the first $k$ components.
One can formulate an even more general inequality by selecting arbitrary $x_i^2y_j^2+x_j^2y_i^2$
instead of all the terms with
$1\leqslant i<j\leqslant k$.

\medskip

Recently, Bong et al  \cite{Bong2018} derived a strong unitary uncertainty relation for any set of unitary operators based
on the positive semi-definiteness of the Gram matrix.
More precisely, let $U_1, \ldots,U_d$ be $d$ unitary operators and $U_0=I$. Their result
says that the positive semidefiniteness of the Gram matrix $G=G(\rho)$ of size $d+1$
with $G_{jk}
=\langle U_j^\dag U_k\rangle=\mathrm{Tr(\rho U_j^{\dagger}U_k)}$ generalizes the UUR. 
In the case of two unitary operators $A$ and $B$,
$\det G(\rho)\geq 0$ turns out to be $\Delta A^2\Delta B^2\geqslant|\langle A^\dag B\rangle-\langle A^\dag\rangle\langle B\rangle|^2$ \cite{Bong2018}, which is exactly the aforementioned (UUR) in Eq.(\ref{prod2}).

We have seen that the lower bound of this UUR is weaker than our Theorem 1.
 In fact for any $2n$ complex numbers $\alpha_i, \beta_i$ \cite{Hardy1952}
 \begin{equation}\label{Cau}
 |\sum_{i=1}^n\alpha_i^*\beta_i|^2\leqslant(\sum_{i=1}^n|\alpha_i||\beta_i|)^2\leqslant\sum\limits_{i,j}|\alpha_i|^2|\beta_j|^2,
 \end{equation}
 where  the second inequality uses the Cauchy-Schwarz inequality.
 It follows from Eq.(\ref{Chain}) that
\begin{align}\label{Compar}
\Delta A^2\Delta B^2&=I_1\geqslant \ldots\geqslant I_k\geq\ldots\geqslant I_N=(\sum_{i=1}^N|\alpha_i||\beta_i|)^2\notag\\
&\geqslant|\sum_{i=1}^N\alpha_i^*\beta_i|^2=|\langle A^\dag B\rangle-\langle A^\dag\rangle\langle B\rangle|^2.
\end{align}
This means that the UUR given in \cite{Bong2018} for two unitary operators
is the weakest bound in this sequence.

As the case of $k=1$ is trivial, we will include this in our statement of the result for simplicity.

\subsection{\label{sec:leve22}Improved UURs}
The symmetric group $S_N$, which acts on the set $\{1, 2, \ldots, N\}$ naturally by permutation, can be used to strengthen the lower bounds of our UURs.  For any two permutations $\pi_1, \pi_2\in S_N$, the induced
action of $S_N\times S_N$ on $I_k$ is given by
\begin{align}\label{permu}
(\pi_1, \pi_2)I_k=&\sum\limits_{1\leqslant i\leqslant N}x_{\pi_1(i)}^2y_{\pi_2(i)}^2\\
&+\sum\limits_{\substack{1\leqslant i<j\leqslant N\\k<j}}(x_{\pi_1(i)}^2y_{\pi_2(j)}^2+x_{\pi_2(j)}^2y_{\pi_1(i)}^2)\notag\\
&+\sum\limits_{1\leqslant i<j \leqslant k}2x_{\pi_1(i)}y_{\pi_2(j)}x_{\pi_2(j)}y_{\pi_1(i)}.\notag
\end{align}
Clearly $I_1$ is stable under the action of $S_N\times S_N$, subsequently
\begin{align}\label{permchain}
I_1\geqslant (\pi_1, \pi_2)I_2\geqslant\ldots\geqslant (\pi_1, \pi_2)I_N.
\end{align}
Optimizing over the symmetric group $S_N$, we obtain the following stronger result.

\
\emph{Theorem} 2.
Let $\rho$ be any quantum state on an $n$-dimensional Hilbert space $H$, $A$ and $B$ two unitary operators on $H$.
One has the following improved unitary uncertainty relations for the product of variances  ($k=1,\ldots,N$)
\begin{equation}
\Delta A^2\Delta B^2\geqslant \max\limits_{\pi_1, \pi_2\in S_N}(\pi_1, \pi_2)I_k,
\end{equation}
where $N=n$ (or $n^2$) if $\rho$ is pure (or mixed), $(\pi_1, \pi_2)I_k$ is
defined in \eqref{permu}, and
the equality holds if and only if $x_{\pi_1(i)}y_{\pi_2(j)}=x_{\pi_2(j)}y_{\pi_1(i)}$ for all $1\leqslant i\neq j\leqslant k$.

\medskip
We remark that the lower bound in Theorem 2
is tighter than that of  Theorem 1, since $\max\limits_{\pi_1, \pi_2\in S_N}(\pi_1, \pi_2)I_k\geqslant I_k$ for any 
$1\leqslant k\leqslant N$. An example is given to show strict strengthening of the bounds (see Example 1 and Fig. \ref{remark}).

\subsection{\label{sec:leve23}Examples}

\emph{Example} 1. Let us consider the pure states $\ket{\psi}=\cos\theta\ket{0}-\sin\theta\ket{d-1}$ on an $d$-dimensional Hilbert space \cite{Bagchi2016}, and $A$, $B$ are the following unitary operators
\begin{align}
A&=\sum_{j=-[\frac d2]}^{[\frac{d-1}{2}]}\omega^j\ket {j}\bra {j}=diag(1, \omega, \omega^2, \ldots, \omega^{d-1}),\notag\\
B&=\sum_{j=-[\frac d2]}^{[\frac{d-1}{2}]}\ket{j+1}\bra{j}=
\begin{pmatrix}
0&1\\
I_{d-1}&0
\end{pmatrix}.
\end{align}
where $\omega=e^{i2\pi /d}$. 
Note that
$AB=\omega BA$ \cite{Massar2008}. 

\begin{figure}[!h]
  \centering
  \includegraphics[width=3.0in]{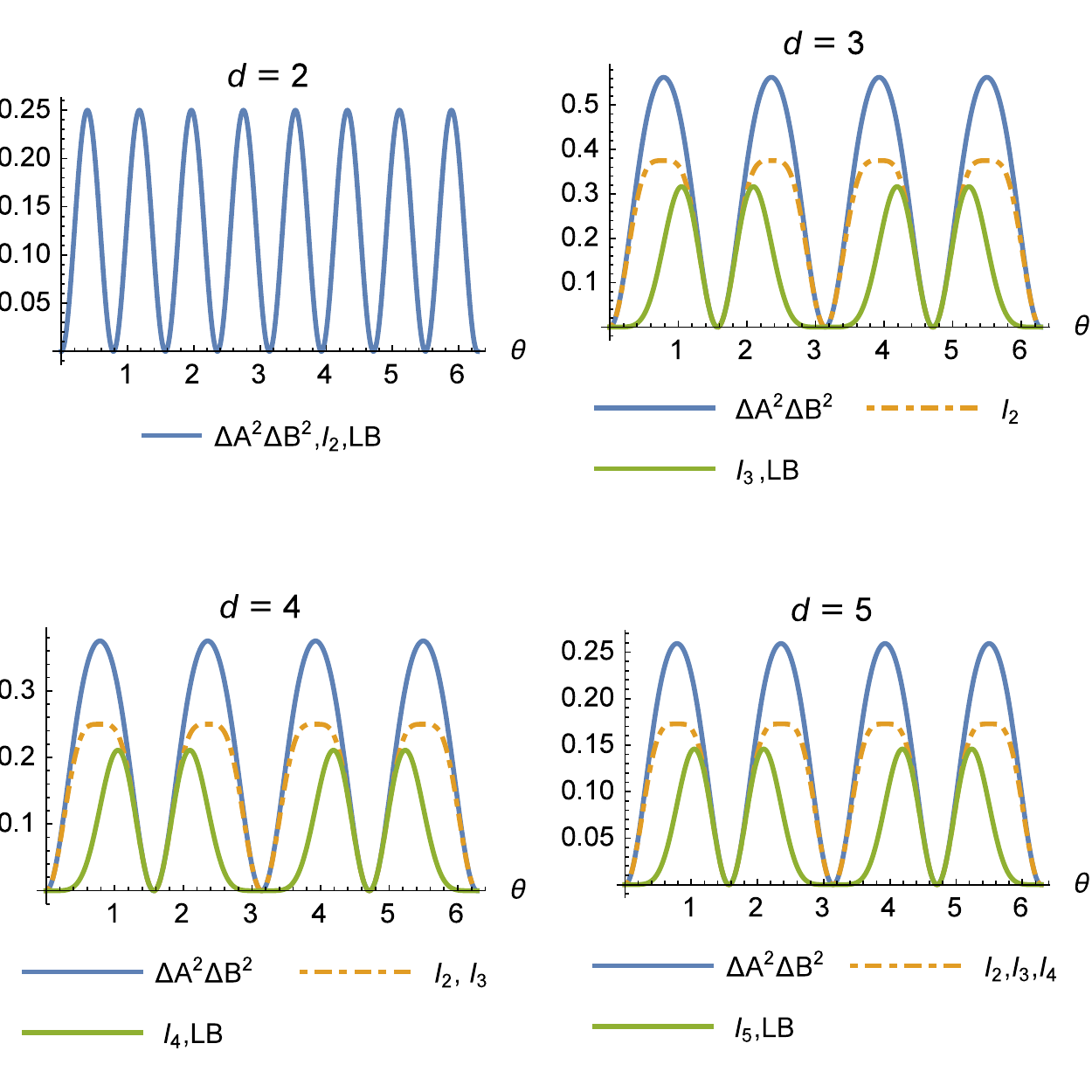}
  \caption{\textbf{Comparison of our bounds with Bong et al's bound for pure state.} 
   The solid blue (upper) and green (lower) curves represent $\Delta A^2\Delta B^2$ and Bong et al's bound LB respectively.
   Our bounds $I_2, I_3$ or $I_4$ are tighter and shown in dashed yellow curves.
   }\label{gra1}
\end{figure}

\emph{Case $d=2$}. In this case
\begin{align}
A=
\begin{pmatrix}
1&0\\
0& -1 
\end{pmatrix},  \quad B 
=\begin{pmatrix}
0&1\\
1&0
\end{pmatrix}.
\end{align}
Both our UUR and Bong et al's are equal to $\Delta A^2\Delta B^2= I_2$ (See Fig.\ref{gra1}).
So we focus on $d=3, 4, 5$, where the UURs are not saturated.

\emph{Case $d=3$}. 
The unitary operators are
\begin{align}
A=diag(1, e^{\frac{2\pi i}{3}}, e^{\frac{4\pi i}{3}}), \quad
B=\begin{pmatrix}
0&1\\
I_2&0
\end{pmatrix}.
\end{align}
their associated real vectors 
$\overrightarrow{X}=(x_1, x_2, x_3)$, $\overrightarrow{Y}=(y_1, y_2, y_3)$ are given by

\begin{align}
x_1&=|(1-e^{-\frac{2\pi i}{3}})\sin^2\theta\cos\theta|, \quad
x_2=0,\notag\\
x_3&=|(1-e^{-\frac{2\pi i}{3}})\sin\theta\cos^2\theta|,
\end{align}

and
\begin{align}
y_1&=|-\sin^3\theta|, \quad
y_2=|\cos\theta|,\notag\\
y_3&=|-\sin^2\theta\cos\theta|,
\end{align}
then $I_2, I_3$ can be fixed and that $\Delta A^2\Delta B^2\geqslant I_2\geqslant I_3=|\langle A^\dag B\rangle-\langle A^\dag\rangle\langle B\rangle|^2$.
Fig.\ref{gra1} shows that our bounds are better than Bong et al's bound.

\emph{Case $d=4, 5$}. The vectors $\overrightarrow{X}$, $\overrightarrow{Y}$ for $d=4, 5$ are respectively as follows.
\begin{align}
&\overrightarrow{X}
=\begin{cases}|(1-e^{\frac{-\pi i}{2}}))\frac{\sin2\theta}2|(|\sin\theta|, 0, 0, |\cos\theta|),\\
|(1-e^{\frac{-2\pi i}{5}})\frac{\sin2\theta}2|(|\sin\theta|, 0, 0, 0, |\cos\theta|)
\end{cases}\\
&\overrightarrow{Y}=
\begin{cases}(|-\sin^3\theta|,|\cos\theta|, 0, |-\sin^2\theta\cos\theta|)\\
(|-\sin^3\theta|,|\cos\theta|, 0,0, |-\sin^2\theta\cos\theta|)
\end{cases}
\end{align}

Then the lower bounds $I_2, I_3, I_4$ (resp. $I_2, I_3, I_4, I_5$) can be computed. It is readily seen that $\Delta A^2\Delta B^2\geqslant I_2= I_3\geqslant I_4=|\langle A^\dag B\rangle-\langle A^\dag\rangle\langle B\rangle|^2$(resp. $\Delta A^2\Delta B^2\geqslant I_2= I_3=I_4\geqslant I_5=|\langle A^\dag B\rangle-\langle A^\dag\rangle\langle B\rangle|^2$). Fig.\ref{gra1} show that
in all these cases, our bounds are better than that of Bong et al.

{Remark. The bounds $I_2, I_3, I_4$ can be further strengthened by Theorem 2. Consider the same qutrit state $\ket{\psi}=\cos\theta\ket{0}-\sin\theta\ket{2}$. Applying the symmetric group $S_3$ as in Eq.(\ref{permu}) it follows that
$\Delta A^2\Delta B^2= \max\limits_{\pi_1, \pi_2\in S_3}(\pi_1, \pi_2)I_2\geqslant\max\limits_{\pi_1, \pi_2\in S_3}(\pi_1, \pi_2)I_3$. Fig.\ref{remark} shows that the new bounds strictly outperform $I_k$.

\begin{figure}[!h]
  \centering
  \includegraphics[width=3.0in]{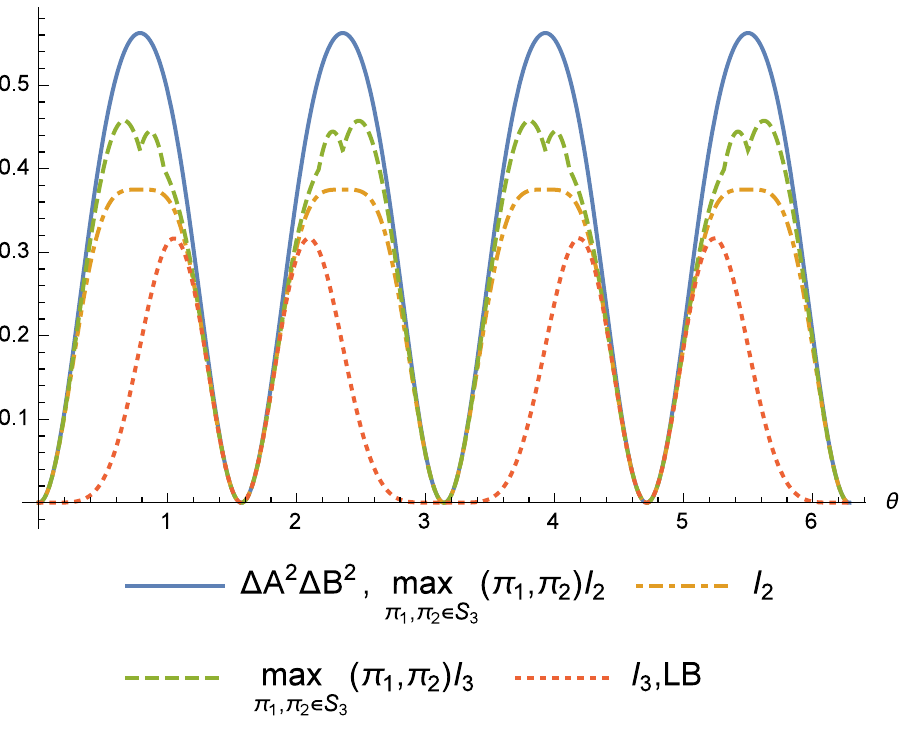}
  \caption{Strengthened bounds vs. the bounds $I_k$ for qutrit pure state.
   The solid blue curve represents $\Delta A^2\Delta B^2$ and $\max\limits_{\pi_1, \pi_2\in S_3}(\pi_1, \pi_2)I_2$. The dashed green curve represents $\max\limits_{\pi_1, \pi_2\in S_3}(\pi_1, \pi_2)I_3$. The dotted dashed and dotted curves represent $I_2$ and LB (or $I_3$) respectively.}\label{remark}
\end{figure}

\emph{Example} 2. Consider the qubit mixed state $\rho=\frac12(I+\vec{r}\cdot\vec{\sigma})$ with $\vec{r}=(\frac13,\frac23\cos\theta,\frac23\sin\theta)$, and $\vec{\sigma}=(\sigma_x,\sigma_y,\sigma_z)$
where $\sigma_x, \sigma_y, \sigma_z$ are the Pauli matrices.

Consider the unitary operators
\begin{align}
A=e^{i\pi\sigma_y/8}&=\begin{pmatrix}\cos\frac\pi8 & \sin\frac\pi8 \\-\sin\frac\pi8 & \cos\frac\pi8\end{pmatrix},\\
B=e^{i\pi\sigma_z/8}&=\begin{pmatrix} e^{i\frac\pi8}&0\\0&e^{-i\frac\pi8}\end{pmatrix},
\end{align}
which correspond to Bloch sphere rotations of $-\pi/4$ about the $y$ axis and $z$ axis respectively.

It is seen that (cf. \ref{appendix}. Appendix A )
\begin{equation}
\ket{\sqrt{\rho}}=\begin{pmatrix}
\frac{\sqrt{3-\sqrt5}(\sqrt5-2\sin\theta)+\sqrt{3+\sqrt5}(\sqrt5+2\sin\theta)}{2\sqrt{30}}\\
\quad\\
-\frac{i(\sqrt{3-\sqrt5}-\sqrt{3+\sqrt5})(-i+2\cos\theta)}{2\sqrt30}\\
\quad\\
\frac{i(\sqrt{3-\sqrt5}-\sqrt{3+\sqrt5})(i+2\cos\theta)}{2\sqrt30}\\
\quad\\
\frac{\sqrt{3+\sqrt5}(\sqrt5-2\sin\theta)+\sqrt{3-\sqrt5}(\sqrt5+2\sin\theta)}{2\sqrt{30}}\\
\end{pmatrix}.
\end{equation}
Then the bounds $I_2,I_3,I_4$ associated with $\rho$ can be computed.
We find that $\Delta A^2\Delta B^2> I_2> I_3> I_4\geq|\langle A^\dag B\rangle-\langle A^\dag\rangle\langle B\rangle|^2$, which is the lower bound
of Bong et al.
Fig.\ref{gra2} shows that our bounds are almost always better than that of Bong et al.
It seems that the bounds $I_k$ are separated for mixed states.

\begin{figure}[!h]
  \centering
  \includegraphics[width=3.5in]{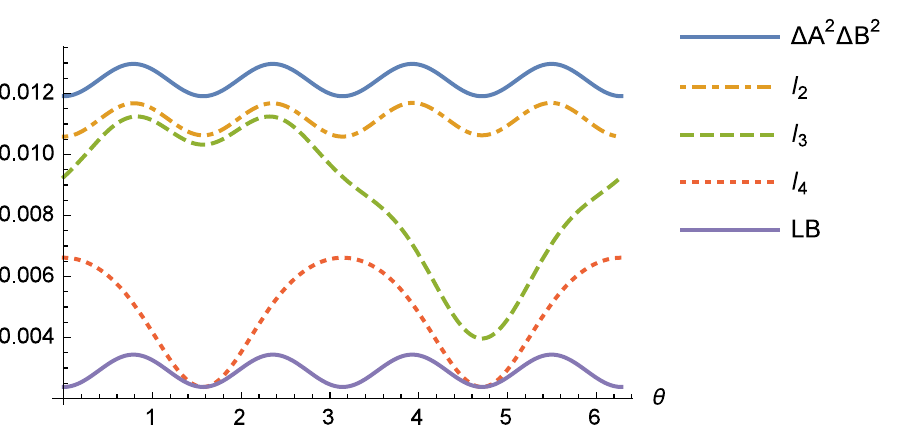}
  \caption{\textbf{Comparison of our bounds with that of Bong et al's for pure state.}
   The solid blue (upper) and purple (lower) curves represent $\Delta A^2\Delta B^2$ and Bong et al's bound LB respectively.
   Our bounds $I_2, I_3$ or $I_4$ are shown in dashed or dotted curves in yellow, green and red respectively.
   }\label{gra2}
\end{figure}

\
\section{\label{sec:leve3}Uncertainty Relations for three unitary operators}
We now study product-form variance-based unitary uncertainty relations for three unitary operators based upon
our UUR for two unitary operators in terms of the quantities $I_k$ in the preceding section.
\subsection{\label{sec:leve31}Main results}
Let $A$,$B$ and $C$  be three unitary operators defined on an $n$-dimensional Hilbert space. By Theorem 1 the 
UURs for the pairs 
$\{A,B\}$, $\{B,C\}$ and $\{A,C\}$ over the quantum state $\rho$ 
are written as $\Delta A^2\Delta B^2\geqslant I_k$, $\Delta B^2\Delta C^2\geqslant J_k$, and $\Delta A^2\Delta C^2\geqslant K_k$
where $I_k, J_k, K_k$ are the quantities $I_k$ defined above \eqref{Chain} for the pairs respectively.
Taking the square root of the product, we have the following result. 

\emph{Corollary} 1. For a fixed quantum state $\rho$ and three unitary operators $A$, $B$ and $C$ on an $n$-dimensional Hilbert space $H$, the product of the variances obeys the following inequalities ($k=2, \ldots, N$)
\begin{equation}
\Delta A^2\Delta B^2\Delta C^2\geqslant (I_kJ_kK_k)^{1/2}
\end{equation}
where $N=n$ (or $n^2$) if $\rho$ is pure (or mixed), $I_k=I_k(A, B)$, $J_k=I_k(A, C)$ and $K_k=I_k(B, C)$.
Here $I_k$ are defined in Sect. \ref{sec:leve21}.

One can also strengthen the bound using the symmetry of $S_N$. Denote
$\max\limits_{\pi_1, \pi_2\in S_N}(\pi_1, \pi_2)I_i$ by $\hat{I}_i$, then
the improved UURs are given in the following corollary.

\emph{Corollary} 2. Let $\rho, A, B, C$ as in Cor. 1.
The strengthened UURs are given by
\begin{equation}
\Delta A^2\Delta B^2\Delta C^2\geqslant (\hat{I}_k\hat{J}_k\hat{K}_k)^{1/2},
\end{equation}
where $\hat{I}_k=\max\limits_{\pi_1, \pi_2\in S_N}(\pi_1, \pi_2)I_k$, $\hat{J}_k=\max\limits_{\pi_1, \pi_2\in S_N}(\pi_1, \pi_2)J_k$ and $\hat{K}_k=\max\limits_{\pi_1, \pi_2\in S_N}(\pi_1, \pi_2)K_k$.

\subsection{\label{sec:leve32}Examples}

For three unitary operators $A, B, C$, Bong et al's UUR is expressed as the positivity of the Gram matrix:
\begin{equation}
\det G(\rho)=\det\begin{pmatrix}
1&\langle A\rangle&\langle B\rangle&\langle C\rangle\\
\langle A^\dag\rangle&1&\langle A^\dag B\rangle&\langle A^\dag C\rangle\\
\langle B^\dag\rangle&\langle B^\dag A\rangle&1&\langle B^\dag C\rangle\\
\langle C^\dag\rangle&\langle C^\dag A\rangle&\langle C^\dag B\rangle&1
\end{pmatrix}
\geqslant0
\end{equation}
which can be rewritten as
\begin{align}\label{grammatrix3}
&\Delta A^2\Delta B^2\Delta C^2\geqslant\Delta A^2|\langle B^\dag C\rangle-\langle B^\dag\rangle\langle C\rangle|^2
\notag\\
&+\Delta B^2|\langle A^\dag C\rangle-\langle A^\dag\rangle\langle C\rangle|^2+\Delta C^2|\langle A^\dag B\rangle-\langle A^\dag\rangle\langle B\rangle|^2\notag\\
&-2Re\{(\langle A^\dag C\rangle-\langle A^\dag\rangle\langle C\rangle)(\langle C^\dag B\rangle-\langle C^\dag\rangle\langle B\rangle)\notag\\
&\quad\quad\quad\quad(\langle B^\dag A\rangle-\langle B^\dag\rangle\langle A\rangle)\},
\end{align}
where $Re$ denotes the real part. 
The right hand side (RHS) will be denoted by LB.
This inequality is saturated for pure state when $n=\dim H\leqslant3$, where the determinant of the Gram matrix vanishes.

Let us compare their result with our bounds in the cases of pure state
($n\geq 4$ ) and mixed state separately.

 \emph{Example} 3. Let 
  $\ket\psi=\frac12\cos{\frac{\theta}2}\ket0+\frac{\sqrt3}{2}\sin{\frac\theta2}\ket1+\frac{1}{2}\sin{\frac\theta2}\ket2+\frac{\sqrt3}{2}\cos{\frac\theta2}\ket3$
and we take three unitary operators: 
\begin{align}
A&=diag(1, e^{i\frac{\pi}{2}}, e^{i\pi},  e^{i\frac{3\pi}{2}}
),
B=\begin{pmatrix}
0&1\\
I_3&0
\end{pmatrix},\notag\\
C&=\begin{pmatrix}
0&1&0&0\\
1&0&0&0\\
0&0&1&0\\
0&0&0&-1
\end{pmatrix}.
\end{align}

Using Corollary 1, the lower bounds $(I_kJ_kK_k)^{1/2}$ ($2\leqslant k\leqslant4$) can be easily calculated and one sees that
they are better than that of Bong et al's in significant regions. See Fig.\ref{gra4} for the comparison.
\begin{figure}[!h]
  \centering
  \includegraphics[width=3.5in]{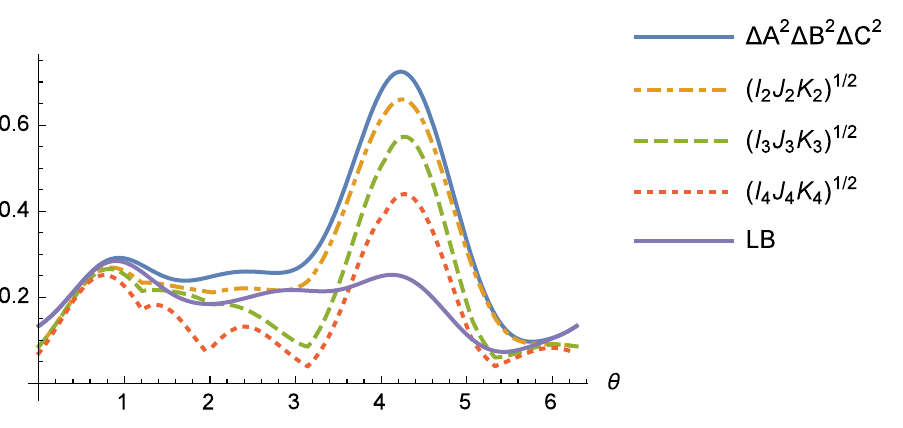}\\
  \caption{\textbf{Comparison of our bounds 
  		 with Bong et al's for pure state.} The solid blue (upper)  and purple (lower) curves are $\Delta A^2\Delta B^2\Delta C^2$ and Bong et al's bound LB.
  	Other three dotted dashed yellow, dashed green, dotted red lines (from top to bottom) represent our bounds $(I_2J_2K_2)^{1/2}$,  $(I_3J_3K_3)^{1/2}$, $(I_4J_4K_4)^{1/2}$ separately.}\label{gra4}
\end{figure}

\emph{Example} 4. Consider the mixed state analyzed in Example 2
and three unitary operators:
\begin{align}
A=e^{i\pi\sigma_y/8}&=\begin{pmatrix}\cos\frac\pi8 & \sin\frac\pi8 \\-\sin\frac\pi8 & \cos\frac\pi8\end{pmatrix},\notag\\
B=e^{i\pi\sigma_z/8}&=\begin{pmatrix} e^{i\frac\pi8}&0\\0&e^{-i\frac\pi8}\end{pmatrix},\notag\\
C=e^{i\pi\sigma_x/8}&=\begin{pmatrix}\cos\frac\pi8 & i\sin\frac\pi8 \\i\sin\frac\pi8 & \cos\frac\pi8\end{pmatrix}.
\end{align}

The vectorized state $\ket{\sqrt\rho}$ was given in Example 2, based on this the uncertainty bound $(I_2J_2K_2)^{1/2}$ can be computed and is
seen to be always tighter than the Bong et al's bound LB
(cf. Fig. (\ref{gra5}).  However, $(I_3J_3K_3)^{1/2}$ and $(I_4J_4K_4)^{1/2}$ are not as good as LB.

\begin{figure}[!h]
  \centering
  \includegraphics[width=3.5in]{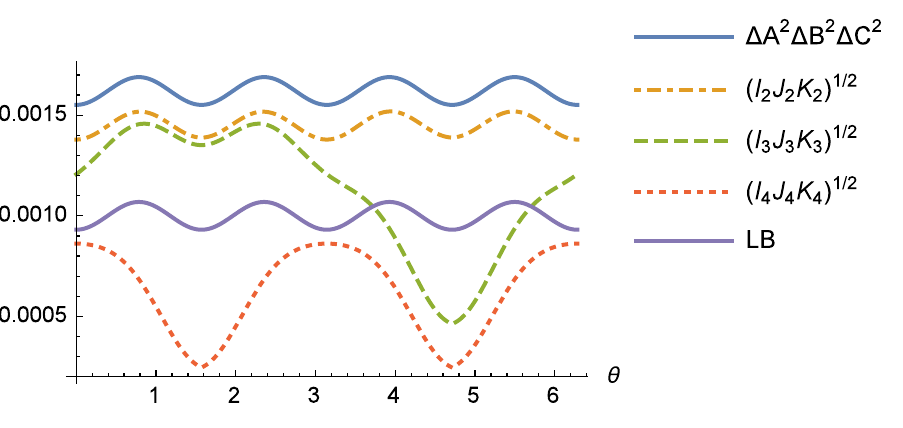}\\
  \caption{\textbf{Comparison of our bounds with Bong et al's bound for mixed state.} The solid blue (upper) and purple ({lower}) curves represent $\Delta A^2\Delta B^2\Delta C^2$ and Bong et al's bound LB. The other three dotted dashed yellow, dashed green, dotted red curves (from top to bottom) are our bounds $(I_2J_2K_2)^{1/2}$,  $(I_3J_3K_3)^{1/2}$, $(I_4J_4K_4)^{1/2}$ respectively. }\label{gra5}
\end{figure}

\

\emph{Example} 5. Consider the mixed qutrit state
$\rho=\frac13(I+\sqrt3\vec{n}\cdot\vec{\lambda})$ \cite{Goyal2016} on $\mathbb C^3$,
where $\vec{\lambda}$
is the $8$-dimensional vector of
the Gell-Mann matrices of rank 3 and

$\vec{n}=(\frac1{\sqrt3}\cos\theta,0,0,0,0,\frac1{\sqrt3}\sin\theta,0,0)$.\quad\quad As a matrix, the density operator $\rho$ takes the following form
\begin{equation}
\rho=\frac13\begin{pmatrix}
1&\cos\theta&0\\
\cos\theta&1&\sin\theta\\
0&\sin\theta&1
\end{pmatrix}.
\end{equation}

The three unitary operators $A,B,C$ are taken as the rotational operators  $R_{Z,\theta_z}, R_{Y,\theta_y}, R_{X,\theta_x}$ with the Euler angles $\theta_z=\frac\pi4,\theta_y=-\frac\pi4,\theta_x=\frac\pi3$ around $Z,Y$ and $X$ axes respectively. i.e.
\begin{align}
R_{Z,\theta_z}&=\begin{pmatrix}
\cos\theta_z&\sin\theta_z&0\\
-\sin\theta_z&\cos\theta_z&0\\
0&0&1
\end{pmatrix},\notag\\
R_{Y,\theta_y}&=\begin{pmatrix}
\cos\theta_y&0&\sin\theta_y\\
0&1&0\\
-\sin\theta_y&0&\cos\theta_y
\end{pmatrix},\notag\\
R_{X,\theta_x}&=\begin{pmatrix}
1&0&0\\
0&\cos\theta_x&-\sin\theta_x\\
0&\sin\theta_x&\cos\theta_x
\end{pmatrix}.
\end{align}

The state $\ket{\sqrt\rho}$ is seen as follows
(cf. \ref{appendix}. Appen. B)

\begin{align*}
\ket{\sqrt\rho}&=(
	\frac{\cos^2\theta}{\sqrt6}+\frac{\sin^2\theta}{\sqrt3},
	\frac{\cos\theta}{\sqrt6},\\
&	\frac{(-2+\sqrt2)\sin{2\theta}}{4\sqrt3},
	\frac{\cos\theta}{\sqrt6},
	\frac{1}{\sqrt6},
	\frac{\sin\theta}{\sqrt6},\\
&	\frac{(-2+\sqrt2)\sin{2\theta}}{4\sqrt3},
	\frac{\sin\theta}{\sqrt6},
	\frac{\cos^2\theta}{\sqrt3}+\frac{\sin^2\theta}{\sqrt6})
\end{align*}

The lower bounds $\{(I_kJ_kK_k)^{1/2}|2\leqslant k\leqslant8\}$ associated with $\rho$ are then calculated and depicted in Fig.\ref{gra6}. The picture shows that our lower bounds  $\{(I_kJ_kK_k)^{1/2}|2\leqslant k\leqslant6\}$ are always tighter than $LB$, Bong et al's bound,  $(I_7J_7K_7)^{1/2}$ and $(I_8J_8K_8)^{1/2}$ are better than LB in some region, and LB is better than $(I_9J_9K_9)^{1/2}$.

\begin{figure}[!h]
  \centering
  \includegraphics[width=3in]{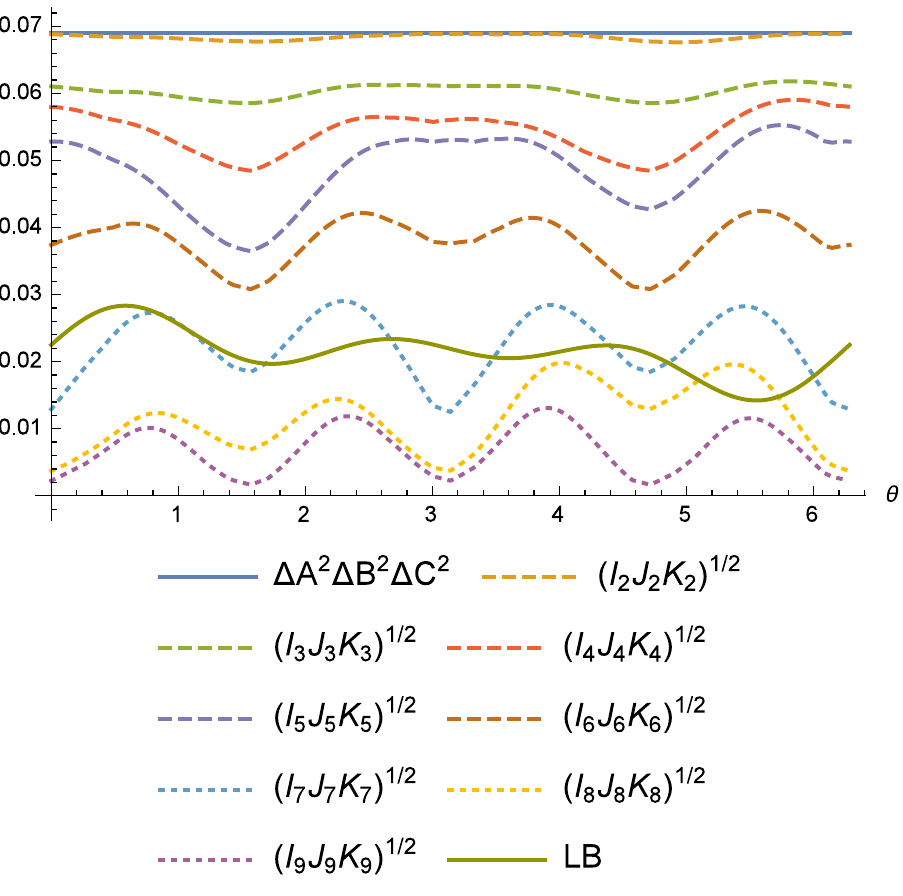}\\
  \caption{\textbf{Comparison of our bounds with Bong et al's bound for qutrit state.} The solid blue (upper) and green (lower) curves represent $\Delta A^2\Delta B^2\Delta C^2$ and Bong et al's bound LB respectively. The other eight dashed or dotted curves (from top to bottom) are the bounds $(I_2J_2K_2)^{1/2}$, \ldots, $(I_9J_9K_9)^{1/2}$. }\label{gra6}
\end{figure}

\section{\label{sec:leve4} Conclusion}
In this paper, we have studied a stronger form of variance-based unitary uncertainty relations (UUR) for two and three operators relative to both pure and mixed quantum states. Our idea is to employ the partial Cauchy-Schwarz inequality to derive a sequence of effective lower bounds $I_k$ for the product of the uncertainties.
Moreover, our bounds $I_k$ can be strengthened by permutation. 

We have also shown that our new uncertainty bounds are tighter
than the recently discovered UUR given by Bong et al using the positivity of the Gram matrix \cite{Bong2018} for two and
multiple unitary operators. In one comparison with Bong et al's bound, two unitary operators related by the discrete Fourier
transform are examined and it was found that our bounds outperform significantly
their lower bounds, which could have potential
implications for signal processing and modular variables.
In another example of three unitary operators,
most of our bounds demonstrated better effects than theirs for arbitrary quantum state and three unitary operators.

\section*{Acknowledgment}
We are grateful to Yunlong Xiao for stimulating discussions and help in this work.
The research is partially supported by National Natural Science Foundation of China grant no. 11531004, Simons Foundation grant no. 523868 and China Scholarship Council.

\section{\label{appendix} Appendices}
\subsection*{\label{appendixa} Appendix A}
The Hermitian matrix $\rho$ is unitarily diagonalizable, so it can be expressed as $\rho=UDU^{\dagger}$ for
a unitary matrix $U$ and a diagonal matrix $D$. 

For the qubit mixed state $\rho=\frac12(I+\vec{r}\cdot\vec{\sigma})$ with $\vec{r}=\{\frac13,\frac23\cos\theta,\frac23\sin\theta\}$ and $\vec{\sigma}=\{\sigma_x,\sigma_y,\sigma_z\}$. The unitary matrix $U=(\frac{v_1}{|v_1|},\frac{v_2}{|v_2|})$, where the orthogonal eigenvectors $u_i, u_2$ are given by
\begin{align*}
v_1&=(-\frac{i \left(2 \sin\theta+\sqrt{5}\right)}{-i+2 \cos\theta},1)^T,\\
v_2&=(\frac{i \left(\sqrt{5}-2\sin\theta\right)}{-i+2\cos\theta},1)^T.
\end{align*}
The diagonal matrix $D$ is determined by the corresponding eigenvalues
and
\begin{equation}
D^{\frac12}=
\begin{pmatrix}
 \sqrt{\frac{1}{6} \left(3+\sqrt{5}\right)} & 0 \\
 0 & \sqrt{\frac{1}{6} \left(3-\sqrt{5}\right)}
\end{pmatrix}.
\end{equation}
Therefore the unique positive semidefinite square root of the Hermitian matrix $\rho$ is given by
\begin{widetext}
\begin{equation}
\sqrt\rho=UD^{\frac12}U^\dag
=\begin{pmatrix}
\frac{\sqrt{3+\sqrt{5}} \left(2 \sin\theta+\sqrt{5}\right)+\sqrt{3-\sqrt{5}} \left(\sqrt{5}-2 \sin\theta\right)}{2 \sqrt{30}} & \frac{i \left(\sqrt{3-\sqrt{5}}-\sqrt{3+\sqrt{5}}\right)(i+2 \cos \theta )}{2 \sqrt{30}} \\
-\frac{i \left(\sqrt{3-\sqrt{5}}-\sqrt{3+\sqrt{5}}\right) (-i+2 \cos \theta )}{2 \sqrt{30}} & \frac{\sqrt{3-\sqrt{5}} \left(2 \sin \theta +\sqrt{5}\right)+\sqrt{3+\sqrt{5}} \left(\sqrt{5}-2 \sin \theta \right)}{2 \sqrt{30}}
\end{pmatrix}.
\end{equation}
\end{widetext}
Consequently, the vectorization $\ket{\sqrt\rho}$ for the mixed state $\rho$ is obtained as a 4-dimensional pure state.

\subsection*{\label{appendixb}Appendix B}
For the qutrit mixed state $\rho=\frac13(I+\sqrt3\vec{n}\cdot\vec{\lambda})$ with $\vec{n}=(\frac1{\sqrt3}\cos\theta,0,0,0,0,\frac1{\sqrt3}\sin\theta,0,0)$ and $\vec{\lambda}=(\lambda_1,\lambda_2,\ldots,\lambda_8)$ is the vector of the Gell-Mann matrices. Using a similar procedure as Appendix A, we diagonalize the matrix $\rho$ as
\begin{equation}
D=U^\dag\rho U=
\begin{pmatrix}
\frac23&0&0\\
0&\frac13&0\\
0&0&0
\end{pmatrix},
\end{equation}
where the unitary matrix $U=(\frac{v_1}{|v_1|},\frac{v_2}{|v_2|},\frac{v_3}{|v_3|})$ is given by the eigenvectors
\begin{align*}
v_1&=(\cot\theta,\csc\theta,1)^T,\\
v_2&=(-\tan\theta,0,1)^T,\\
v_3&=(\cot\theta,-\csc\theta,1)^T.
\end{align*}
Then the unique semidefinite square root of matrix $\rho$ is
\begin{equation}
\sqrt\rho=UD^{\frac12}U^=\begin{pmatrix}
 \frac{\cos^2\theta}{\sqrt6}+\frac{\sin^2\theta}{\sqrt3} & \frac{\cos\theta}{\sqrt6} & \frac{(-2+\sqrt2)\sin{2\theta}}{4\sqrt3} \\
 \quad\\
 \frac{\cos\theta}{\sqrt6} & \frac{1}{\sqrt6} & \frac{\sin\theta}{\sqrt6} \\
 \quad\\
 \frac{(-2+\sqrt2)\sin{2\theta}}{4\sqrt3} & \frac{\sin\theta}{\sqrt6} & \frac{\cos^2\theta}{\sqrt3}+\frac{\sin^2\theta}{\sqrt6} \\
\end{pmatrix}.
\end{equation}
By stacking columns of the matrix $\sqrt{\rho}$ on top of one another, we have the pure state $\ket{\sqrt\rho}$ on the 9-dimensional Hilbert space.

\subsection*{\label{appendixc}Appendix C}
To highlight our method, we further consider the strengthened UURs for four unitary operators.

Let $A$, $B$, $C$ and $D$  be four unitary operators on an $n$-dimensional Hilbert space, the product form of variance-based unitary uncertainty relations with two pairs of unitary operators $\{A,B\}$ and $\{C,D\}$ in quantum state $\ket{\psi}$ can be written as $\Delta A^2\Delta B^2\geqslant I_k$, $\Delta C^2\Delta D^2\geqslant J_k$ respectively.

Therefore UURs for four unitary operators is then given as follows:
\begin{equation}\label{fourrelation}
\Delta A^2\Delta B^2\Delta C^2\Delta D^2\geqslant I_kJ_k,
\end{equation}
with $2\leqslant k\leqslant N$.

Though the above seems to be a trivial step beyond the
case of two unitary operators, 
it still outperforms Bong et al's bound in many situations. 

\emph{Example} 6. Let us consider the pure state  $\ket{\psi}=\cos\theta\ket{0}+\frac12\sin\theta\ket{1}+\frac{\sqrt3}{2}\sin\theta\ket{4}$ on 5-dimensional Hilbert space, and take four unitary operators $A$, $B$, $C$ and $D$ as follows.
\begin{align}
A&=diag(e^{-\frac{4\pi i}{5}}, e^{-\frac{2\pi i}{5}}, 1, e^{\frac{2\pi i}{5}}, e^{\frac{4\pi i}{5}}),\notag\\
B&=diag(e^{\frac{4\pi i}{5}}, e^{\frac{2\pi i}{5}}, 1, e^{-\frac{2\pi i}{5}}, e^{-\frac{4\pi i}{5}} ),\\
C&=\begin{pmatrix}
0&1\\
I_4&0
\end{pmatrix},
D=i\begin{pmatrix}
0&1\\
I_4&0
\end{pmatrix}.\notag
\end{align}
It is not difficult to check that $\Delta A^2\Delta B^2\Delta C^2\Delta D^2= I_kJ_k$ with $2\leqslant k\leqslant5$ in our UURs due to its saturated conditions.

For four unitary operators, Bong et al's UUR is
\begin{align}
\det G(\rho)&=\det\begin{pmatrix}
1&\langle A\rangle&\langle B\rangle&\langle C\rangle&\langle D\rangle\\
\langle A^\dag\rangle&1&\langle A^\dag B\rangle&\langle A^\dag C\rangle&\langle A^\dag D\rangle\\
\langle B^\dag\rangle&\langle B^\dag A\rangle&1&\langle B^\dag C\rangle&\langle B^\dag D\rangle\\
\langle C^\dag\rangle&\langle C^\dag A\rangle&\langle C^\dag B\rangle&1&\langle C^\dag D\rangle\\
\langle D^\dag\rangle&\langle D^\dag A\rangle&\langle D^\dag B\rangle&\langle D^\dag C\rangle&1
\end{pmatrix}\notag\\
&\geqslant0
\end{align}
It is complicated and cumbersome to simplify the above into a form of $\Delta A^2\Delta B^2\Delta C^2\Delta D^2\geqslant M$, the uncertainty lower bound. So we simply sketch  $\det G(\ket{\psi})$ in Fig.\ref{gra7}. We find that the determinant $G(\ket{\psi})$ vanishes only when $\{\theta=n\pi|n\in\mathbb{Z}\}$, i.e., when the uncertainty relation is saturated.

This means that our bound is tighter than Bong et al's bound in the whole range except at the points $n\pi$. Given the complexity of straightening out the
product of the variances from $\det G(
\rho)$ as required from Bong et al's method, our procedure is simpler and provides direct lower bounds in this case.

\begin{figure}[!h]
  \centering
  \includegraphics[width=3in]{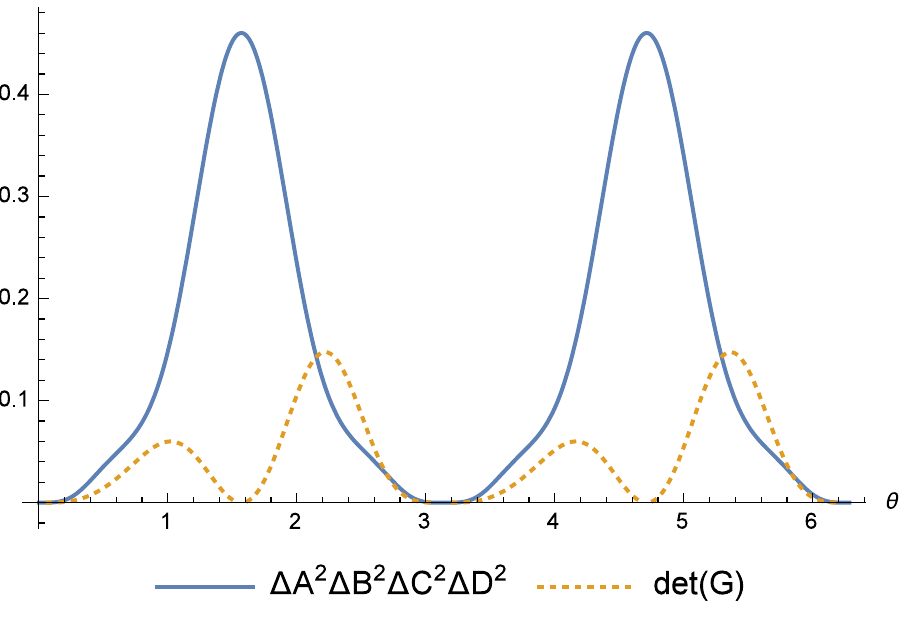}\\
  \caption{The solid blue curve represents $\Delta A^2\Delta B^2\Delta C^2\Delta D^2$, the dotted orange line denotes $\det G(\ket{\psi})$. }\label{gra7}
\end{figure}

\bibliographystyle{amsalpha}

\end{document}